\documentclass[aps,prb,twocolumn,amsmath,amssymb,showpacs,superscriptaddress,notitlepage,longbibliography]{revtex4-2}
\usepackage{graphicx}
\usepackage{subfigure}
\usepackage{dcolumn}
\usepackage{bm}
\usepackage{amsfonts}
\usepackage{amssymb}
\usepackage{amsmath}
\usepackage{color}
\usepackage[colorlinks=true,breaklinks=true,linkcolor=blue,anchorcolor=red,citecolor=blue,urlcolor=blue]{hyperref}
\allowdisplaybreaks[3]

\def\bea{\begin{eqnarray}}
\def\eea{\end{eqnarray}}

\def\nn{\nonumber}
\def\Eq#1{Eq.~(\ref{#1})}

\def\Fig#1{Fig.~\ref{#1}}

\def\xk#1{\left(#1\right)}
\def\zk#1{\left[#1\right]}

\def\sgn{{\rm sgn}}

\def\pa{\partial}
\newcommand{\s}{{\sigma}}

\newcommand{\ga}{\gamma}

\newcommand{\la}{\lambda}
\newcommand{\La}{\Lambda}

\renewcommand{\v}[1]{{\bf #1}}

\begin{document}

\title{Reply to Comment on \href{https://journals.aps.org/prl/abstract/10.1103/PhysRevLett.127.176601}{Phys. Rev. Lett. 127, 176601 (2021)} by Lee and Yang}

\author{Peng-Lu Zhao}
\affiliation{Shenzhen Institute for Quantum Science and Engineering and Department of Physics, Southern University of Science and Technology (SUSTech), Shenzhen 518055, China}
\affiliation{Shenzhen Key Laboratory of Quantum Science and Engineering, Shenzhen 518055, China}

\author{Xiao-Bin Qiang}
\affiliation{Shenzhen Institute for Quantum Science and Engineering and Department of Physics, Southern University of Science and Technology (SUSTech), Shenzhen 518055, China}
\affiliation{Shenzhen Key Laboratory of Quantum Science and Engineering, Shenzhen 518055, China}

\author{Hai-Zhou Lu}
\email{Corresponding author: luhz@sustech.edu.cn}
\affiliation{Shenzhen Institute for Quantum Science and Engineering and Department of Physics, Southern University of Science and Technology (SUSTech), Shenzhen 518055, China}
\affiliation{Shenzhen Key Laboratory of Quantum Science and Engineering, Shenzhen 518055, China}

\author{X. C. Xie}
\affiliation{International Center for Quantum Materials, School of Physics, Peking University, Beijing 100871, China}
\affiliation{CAS Center for Excellence in Topological Quantum Computation, University of Chinese Academy of Sciences, Beijing 100190, China}
\affiliation{Beijing Academy of Quantum Information Sciences, West Building 3, No. 10, Xibeiwang East Road, Haidian District, Beijing 100193, China}

\date{\today}

\begin{abstract}
In this Reply, we respond to the comments in \cite{Lee2023} on our \href{https://journals.aps.org/prl/abstract/10.1103/PhysRevLett.127.176601}{Phys. Rev. Lett. 127, 176601 (2021)} “Coulomb instabilities of a three-Dimensional higher-order topological insulator". We show the surface gap given in \cite{Lee2023} is different from the expression derived by using the well-accepted approach and becomes divergent and singular at lower energies, thus is not suitable for depicting the phase transition from the 2nd-order to 1st-order topological insulator. We further show that a correct surface gap can describe the phase transition if the RG scheme treats the bulk gap as starting point. We justify our criteria in \cite{Zhaopl21PRL} for both the transitions from 2nd-order topological insulator to 1st-order topological insulator and normal insulator. 
\end{abstract}

\maketitle

First of all, the results in \cite{Lee2023} have verified the correctness of our calculations in \cite{Zhaopl21PRL} and shown the significance and broad interest of this topic. The comments in \cite{Lee2023} question our criterion for the transition from the 2nd- to 1st-order topological insulator. A new criterion, the surface gap, is raised in \cite{Lee2023}, because the 2nd-order topological insulator has surface gaps while the 1st-order topological insulator has not. 

{\color{blue} \emph{The divergent and singular surface gap is not suitable for describing topological phase transition.} } We have two concerns on the surface gap in \cite{Lee2023}. 

First, its expression \cite{Lee2023}
\bea
m_{\text{surf}}=m D/B\label{Eqgap}
\eea
is different from the one 
\bea
m'_{\text{surf}}=m D/\sqrt{D^2+B^2}\label{Eqsurfgap}
\eea
derived by using the well-accepted approach \cite{Shanwy10NJP,Luhz10PRB} (see details in Appendix). 
Here, $m$, $D$, and $B$ are model parameters in the Hamiltonian that we use in \cite{Zhaopl21PRL}
\begin{eqnarray}\label{Eq:HOTI}
H&=& H_{\text{TI} }+H_{\text{D}}, \nonumber\\
H_{\text{TI} }&=&\xk{m+B\v k^2}\ga_0+v\bm{\ga} \cdot \v k,\nonumber\\
H_{\text{D}}&=&-D\xk{k_x^2-k_y^2}\ga_5, 
\end{eqnarray}
where $H_\mathrm{TI}$ describes a 1st-order topological insulator and the $D$ term breaks time-reversal symmetry to open side-surface gaps on $H_\mathrm{TI}$, to generate a 2nd-order topological insulator. Without the $D$ term, the model reduces to the 1st-order TI and there is no surface gap, so the vanishing $D$ is a better criterion for describing the transition from the 2nd- to 1st-order topological insulator.

\begin{figure}[htb]
\centering\vspace{-0.1cm}\hspace{-0.2cm}
\vspace{-0.1cm}
\includegraphics[width=0.867\columnwidth]{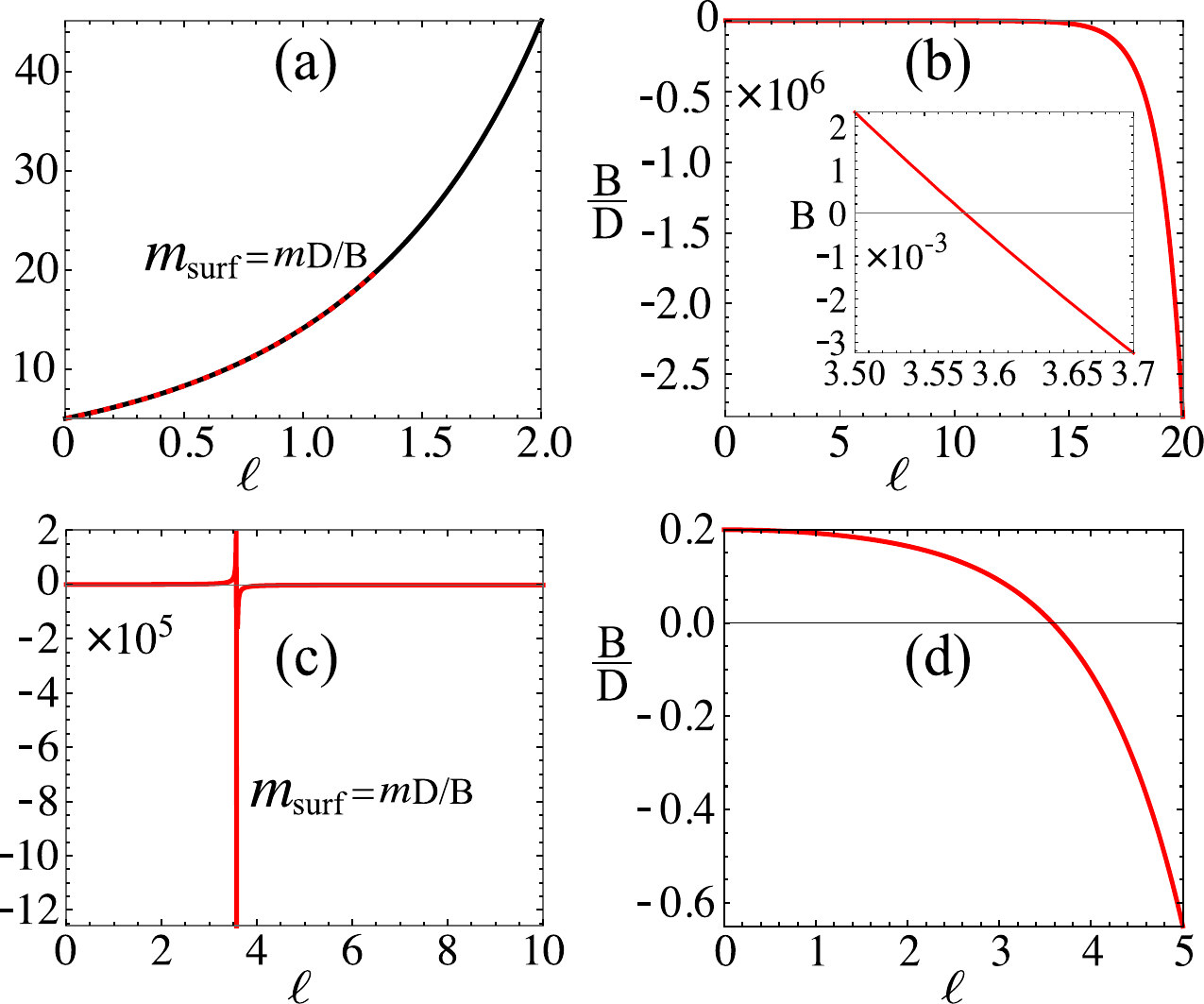}
\caption{(a) The surface gap $m_{\text{surf}}\equiv mD/B$ defined in \cite{Lee2023} as a function of $\ell$. Larger $\ell$ means lower energy scale. Our results (black solid) recover those (red dashed) from Fig. 1(a) in \cite{Lee2023}. (b) $B/D$ decreases with increasing $\ell$ without bound. Inset: $B=0$ around $\ell_{c}\approx 3.6$, leading to the singular $m_{\text{surf}}$.  (c) $m_{\text{surf}}$ becomes singular around $\ell_{c}\approx 3.6$. (d) For $\ell<5$, $B/D<1$, violating the assumption of $D\ll B$ made in the perturbation calculation of the surface gap in \cite{Lee2023}. For all diagrams, the initial values, namely $m_0=B_{\perp}^0=1$, $B_z^0=0.5$, $\alpha_0=\gamma_0^2=0.1$, and $D_0=5$, are identical to the values specified in \cite{Lee2023}. }\label{Figflow}
\end{figure}

Second, the flow of the surface gap $m_{\mathrm{surf}}$ is shown in \cite{Lee2023}, but only up to $\ell=1.3$, far away from the low-energy thermodynamic limit (i.e., $\ell\rightarrow \infty$). \Fig{Figflow}(a) shows that our results recover $m_{\mathrm{surf}}$ in \cite{Lee2023} from $\ell=0$ through 1.3. Strikingly, as we further increase $\ell$, we notice a singular behavior around $\ell\sim 3.6$ for $m_{\mathrm{surf}}$, as depicted in \Fig{Figflow}(c). As shown in the inset diagram of Fig. \ref{Figflow}(b), this singular behavior of $m_{\mathrm{surf}}$ is caused by the vanishing of $B$ at $\ell\sim 3.6$, so does the increased behavior of $m_{\mathrm{surf}}$. Therefore, the claimed increased surface gap in \cite{Lee2023} is misleading. This singular behavior shows that $m_{\mathrm{surf}}$ of \Eq{Eqgap} is not suitable for describing the phase transition.

Moreover, further analysis of how the authors in \cite{Lee2023} obtained \Eq{Eqgap} revealed that their perturbation calculation assumed $D\ll B$, which is not satisfied in the region where the increased gap is described by \Eq{Eqgap}. \Fig{Figflow}(d) shows that for $\ell<5$, $D>B$, and even more critically, $\ell \sim 3.6$,  $D\gg B$, which conflicts with the use of \Eq{Eqgap} to describe the surface gap.

{\color{blue}\emph{The surface gap vanishes if RG is properly handled}}--Now we show that the surface gap vanishes in the low-energy limit as a result of the Coulomb interaction, if the RG flow of the bulk gap is properly handled.  Our RG calculations show that $B$ and $D$ always flow to zero and bulk gap $m$ diverges as $\ell\rightarrow \infty$. 
The behaviors can be understood from the RG scheme,  in which the linear-$k$ term is the free-field Hamiltonian, while the $k^0$ term ($m$) and the $k^2$ terms ($B$ and $D$) are treated as perturbations. Integrating out high-energy modes is equivalent to approaching the $k=0 $ point,  resulting in two consequences. 
First, the $k^2$-related $B$ and $D$ become less important and thus can be ignored as $k\rightarrow 0$ (i.e., $\ell\rightarrow \infty$). Second, the $k^0$ term $m$ becomes more important thus diverges as $\ell\rightarrow \infty$. However, the bulk gap $m$ must be finite in a realistic system. Once $m$ flows to comparable with the cutoff $\La \sim 1/a$ of the system, it is not proper to view the linear-$k$ term (i.e., $v|\mathbf{k}|$ in Eq. (\ref{Eq:HOTI})) as the free-field Hamiltonian. Instead, one needs to view the $m$ term as the starting Hamiltonian and all other terms as perturbations \cite{Peskinb}.  As a result, with increasing $\ell$, $m$ never flows, but $B$ and $D$ still decrease.  According to \Eq{Eqsurfgap}, the gap in the low-energy limit becomes
\bea
m'_{\text{surf}}\big|_{\ell \rightarrow \infty}=\La /\sqrt{1+(B/D)^2|_{\ell \rightarrow \infty }}\rightarrow 0,
\eea
which shows that the surface gap does not change sign when the transition from the 2nd- to 1st-order topological insulator happens, instead, it just vanishes, if the divergent bulk gap is properly handled. More importantly, the transition is controlled by the unbounded increase of $|B/D|$ (see \Fig{Figflow}(b)), where both $B$ and $D$ are of the same $k^2$ order, so their tree-level scaling are cancelled, indicating that the Coulomb interaction is truly behind the transition.

{\color{blue}\emph{A vanishing $B$ does not influence the topology }}--Although $B$ also flows to zero, the vanishing $B$ does not indicates the transition from the 1st-order topological insulator to normal insulator. To illustrate this, we note that the topology of the 1st-order topological insulator is described by the Pfaffian  \cite{Shen11spin}
\bea
P(\v k)=i \frac{m-B k^{2}}{\sqrt{\left(m-B k^{2}\right)^{2}+v^{2}  k^{2}}}
\eea
at $k=0$ and $k\rightarrow \infty$, specifically, 
\begin{eqnarray}
P(k=0)=i \sgn(m), \ \ P(k\rightarrow \infty)= -i \sgn(B).
\end{eqnarray}
In other words, $m$ only matters at $k=0$ 
and $B$ only matters as $k\rightarrow\infty$, respectively.  
Moreover, by approaching to $k= 0$, the RG scheme cannot change $B$ at $k\rightarrow \infty$, so it is safe to use only $m$ to describe the change of the topology of the 1st-order topological insulator.

{\color{blue}\emph{Conclusion}}-- 
We have justified our criteria \cite{Zhaopl21PRL} for both of the transitions from 2nd-order topological insulator to 1st-order topological insulator and normal insulator. 
Nevertheless, we would like to emphasize that the transition from a second-order to a first-order topological insulator is an emergent phenomenon, which is supposed to happen only in the lowest-energy limit. The transition from the 2nd-order topological insulator to normal insulator needs a critical strength of the Coulomb interaction. More importantly, our RG calculations are at zero temperature. Finite temperatures may stabilize the higher-order topological insulator, which is a topic that deserves further exploration.

This work was supported by the National Key R$\&$D Program of China (Grant No. 2022YFA1403700), the Innovation Program for Quantum Science and Technology (Grant No. 2021ZD0302400), the National Natural Science Foundation of China (Grant No. 11925402), Guang-dong province (Grants No. 2020KCXTD001 and No. 2016ZT06D348), the Science, Technology and Innova- tion Commission of Shenzhen Municipality (Grants No. ZDSYS20170303165926217, No. JAY20170412152620376, and No. KYTDPT20181011104202253). The numerical calculations were supported by Center for Computational Science and Engineering of SUSTech.


\appendix

\section{Gap of the surface states}

We show the detailed calculations of the surface gap in the $(100)$ plain, we focus on the gap instead of the whole dispersion of the surface states, and thus set $k_y=k_z=0$. By replacing $k_x$ with $-i\pa_x$, The Hamiltonian becomes
\bea
H&=&\xk{m+B\pa_x^2}\tau_z\s_0-iv\pa_x \tau_x\s_x +D\pa_x^2\tau_y \s_0
\eea
We directly solve the surface models by the method used in \cite{Shanwy10NJP,Luhz10PRB}. First, we take a trial wave-function
\bea
\psi=\psi_{\la} e^{\la x},
\eea
the secular equation for energy $E$ is given by
\bea
\det|H(k_x=-i \la)-E|=0,
\eea
which gives four solutions of $\la$,
\bea
\la_{\alpha}=\pm \sqrt{\frac{X+(-1)^{\alpha}\sqrt{X^2-4Y}}{2 \left(B^2+D^2\right)} }, \label{Eqlambdai}
\eea
where $X=v^2-2 B m$, and $Y=\xk{B^2+D^2}\left(m^2-E^2\right)$. For a $\beta \la_{\alpha}$, the corresponding two eigenstates for $H(k_x=-i \la)$ with eigenvalue $(-1)^{\alpha}E$ are found to be
\bea
\psi_{\alpha \beta 1}&=&\zk{i v \beta \lambda_{\alpha},i D \lambda_{\alpha}^2,0,Z_{\alpha} }^{T},
\nn\\
\psi_{\alpha \beta 2}&=&\zk{i D \lambda_{\alpha}^2,i v \beta \lambda_{\alpha},Z_{\alpha}, 0}^{T} 
\eea
where $Z_{\alpha}=B \lambda_{\alpha}^2+m-E$, $T$ represents the transpose. The general wave functions are given by 
\bea
\Psi=\sum_{\alpha=1,2}\sum_{\beta=\pm}\sum_{\gamma=1,2}C_{\alpha \beta \ga}\psi_{\alpha \beta \ga}e^{\beta \la_{\alpha} x}.
\eea
We consider a semi-infinite boundary conditions:
\bea
\Psi\xk{x=0}=0 \quad \text{and} \quad \Psi\xk{x=+\infty}=0.
\eea
$\Psi\xk{x=+\infty}$ requires that only terms with $\beta=-$ are kept because we assumed that $\la_{\alpha}$ have positive real parts. For $\beta=+$, to have nonzero $C_{\alpha \beta}$ of $\Psi\xk{x=0}=0$, the secular equation leads to 
\bea
v^2\xk{m-E-B\la_1 \la_2}^2=D^2\xk{m-E}^2\xk{\la_1+\la_2}^2,\label{Eqcondition}
\eea
where the condition $\la_1\neq \la_2$ is assumed. Substituting \Eq{Eqcondition} into \Eq{Eqlambdai}, the dispersion of surface state are given by
\bea
E_{\pm}=\pm m D/\sqrt{D^2+B^2}.
\eea

\bibliography{HOTICoul0803}

\end{document}